# A Robust High-temperature Multiferroic Device with Tunable Topological Hall Effect

Zhi Shiuh Lim[1], Changjian Li[2,#], Ping Yang[3], A. Ariando[1,*]

[1]Department of Physics, Faculty of Science, 2 Science Drive 3, National University of Singapore, Singapore 117551

[2]Department of Materials Science and Engineering, Engineering College 442, Southern University of Science and Technology, Shenzhen City, Guangdong Province, China 518055

[3]Singapore Synchrotron Light Source, 5 Research Link, National University of Singapore, Singapore 117603

Emails of corresponding authors: [#]licj@sustech.edu.cn; [*]ariando@nus.edu.sg

**Abstract:**

We fabricated Pt/La$_{0.5}$Ba$_{0.5}$MnO$_3$ ultrathin films with integration to the PbZr$_{0.2}$Ti$_{0.8}$O$_3$ ferroelectric. Strong Topological Hall Effect can be measured across a wide temperature range, which can be turned on and off, corresponding to the ferroelectric polarization switching driven by short voltage pulses, indicating the creation and annihilation of magnetic Skyrmions. However, its magnetoelectric coupling polarity was the opposite of the earlier consensus, which involves a phase transition between ferromagnetic and A-type antiferromagnetic in manganites due to electrostatic doping into the $e_g$-orbitals by the ferroelectric bound charges. We propose a novel yet complementary picture where the transition occurs between C-type antiferromagnetic and ferromagnetic. The distinction between the earlier consensus and our picture lies in the tetragonal distortion of the manganite unit cell, resulting in an opposite magnetoelectric coupling polarity. We support the perceived relationship between the phase transition and tuning of the Topological Hall Effect by micromagnetic simulations.

**Main text:**

Statistical projection shows that by 2025, the information technology sector alone would consume up to 20% of the total electricity produced worldwide and contribute up to 5.5% of the world's carbon emissions. Hence, reducing energy consumption has become the primary concern in the development of next-generation computational and data storage devices. The spintronic research field largely aims at building non-volatile magnetic random-access memory (MRAM) to replace the former volatile static (SRAM) and dynamic (DRAM) variants so that the possibility of memory loss and their requirement of periodic refreshing can be avoided. In MRAM, controlling the state of a ferromagnetic domain memory cell by electric current via the spin-transfer torque (STT) or spin-orbit torque (SOT) mechanism is of paramount importance for device miniaturization. Yet, the high current densities involved in switching collinear magnetic domain (~$10^{10}$-$10^{11}$ A/m$^2$) still suffer a significant bottleneck[1]. Then, interests in building racetrack memories based on magnetic Skyrmions in replacement of collinear domains have surfaced[2, 3, 4], stemming from the discovery where a much lower threshold current density ($J_C$~$10^6$ A/m$^2$) is needed for initiating Skyrmions' motion, thanks to their topological property playing the crucial role in evading domain-pinning by impurities[5, 6]. On the other hand, the magnetoelectric coupling, i.e. the manipulation of magnetic properties by electric field, or the narrower non-volatile subset – multiferroicity, are also actively pursued[7] since the usage of electric current and Joule heating problems can be eliminated altogether.

Thus far, among non-multiferroic materials, various magnetic properties such as coercive fields[8], Anomalous Hall Effect (AHE)[9, 10] and anisotropy[11] have shown decent responses to electric field. Their relevant atomistic mechanisms include changes in carrier concentration of a magnetic semiconductor affecting its *p-d* orbitals exchange interaction[9, 12], changes in Fermi level and 3d-orbital occupancies affecting the spin-obit coupling (SOC) strength in the Fe3d-O2p hybridized interfacial electronic structure[13], or the spin-dependent screening length of electric field[14]. Whereas intrinsic bulk multiferroic materials such as the Bi-based perovskites (BiFeO$_3$, BiMnO$_3$) have led to the demonstration of tuning exchange bias with an adjacent ferromagnet[15] and a 4-state tunnelling junction[16]. Besides, the [111]-oriented electric polarization of the bloch-type Skyrmion-host Cu$_2$OSeO$_3$ is directly related to its magnetic texture phases and thus can be exploited to create an extended metastable Skyrmion phase[17, 18]. Most recently, the Kagome lattice frustrated antiferromagnet $Mn_3Sn$ exhibited large piezomagnetic tunings of AHE induced by strains[19]. Besides, a multiferroic two-dimensional electron gas composed of $LaAlO_3$/$EuTiO_3$/Ca:$SrTiO_3$ also displayed clear responses of AHE and magnetoresistance to electric field[20].

Although choices of bulk multiferroic materials are scarce, multiferroic interfaces involving a conventional ferroelectric and an ultrathin ferromagnet can be relatively easier to make[21, 22, 23]. This is especially true if the weak ferromagnetic properties of ultrathin films are appreciated, in combination with Dzyalloshinskii-Moriya Interaction (DMI), as in Neel-type Skyrmions. Hence, attempts on strain-induced tuning of metallic Néel-type Skyrmions by a large-$d_{33}$ ferroelectric such as the $Pb_{0.64}(Mg_{1/3}Nb_{2/3})O_3$-$Pb_{0.36}TiO_3$ (PMN-PT) are progressively seen[24, 25]. In contrast to earlier efforts, some recent work involved tuning of DMI strength and changes in Skyrmion's helicity has been observed[26, 27]. In this work, we fabricated a Pt/$La_{0.5}Ba_{0.5}MnO_3$/$PbZr_{0.2}Ti_{0.8}O_3$/$SrRuO_3$ (Pt/LBMO/PZT/SRO) heterostructure grown on $SrTiO_3$(001) substrate, with device finish. We observed reversible and non-volatile manipulation of the Topological Hall Effect (THE) at near room temperatures, indicating the creation and annihilation of magnetic Skyrmions upon switching the PZT ferroelectric polarization.

Previously, room-temperature magnetic Skyrmions and other topological textures have been imaged in several manganites stabilized by dipolar interaction[28, 29] or strain gradient-induced DMI[30]; while Pt is the obvious DMI contributor in our case. Fig. 1a shows the total AHE of Pt(2nm)/LBMO(6uc) arising from Spin-Hall-related magnetic proximity effect (MPE), spanning across a wide temperature range. It typically follows a $\rho_{AHE} \propto (T-T_c)^{-\alpha}$ power-law divergence trend with temperature[31], hence reaching its strongest at the Curie temperature ($T_C$). Albeit Skyrmion imaging has not been achieved in such Pt-oxide systems[32, 33, 34] due to weak magnetization below typical limits of imaging instruments, the hump-shape signal has been well-accepted as the topological *subset* of AHE (i.e.: THE), indicating the presence of Skyrmions. Further clarification: THE arises from mobile electrons interacting with Skyrmions by spin-moment exchange ($\hat{\sigma} \cdot \hat{m}$) exchange which does not require SOC in electron deflection[35, 36, 37, 38], unlike the Karplus-Luttinger (KL)[39] *subset* of AHE for collinear magnetizations. Hence, THE produces the hump-shape indicating the formation of peak Skyrmion density with a net solid angle ($\Omega$) subtended by a chiral arrangement of neighbouring magnetic moments $\hat{m}_i \cdot (\hat{m}_j \times \hat{m}_k)$ before gradual saturation into collinear. Although the earlier debate is still mathematically applicable here, which may discredit the claim of THE by suggesting that the humps can be reproduced artificially by two sigmoidal KL-AHE of opposite signs[40, 41], it can be ruled out by a magnetic field θ-angle rotation Hall measurement scheme from out-of-plane to in-plane[42] (Fig. 1b). The scheme shows that the hump-peak field is always stationary with θ-variation and consistent to Skyrmion-lattice behaviour. However, the hump peak fields ($H_{peak}$) arising from two overlapping KL-AHE are expected to diverge following a $1/\cos^\gamma(\theta)$ trend, where γ>0 is a scaling exponent. In Supplementary Fig. 1, we show that the non-divergent behaviour of $H_{peak}$ is similar to the Pt(2nm)/Tm$_3$Fe$_5$O$_{12}$(3uc) discussed earlier[42], i.e., consistent with the Ginzburg-Landau's framework where Skyrmions (triple-*q* spin-wave superposition) will be elongated and annihilated by the in-plane magnetic field component into single-*q* stripes and can be proven by a MUMAX$^3$ micromagnetic simulation mapping of two-dimensional topological charge density (TCD) = $\hat{m} \cdot \left(\frac{d\hat{m}}{dx} \times \frac{d\hat{m}}{dy}\right)$ versus θ-angle and total magnetic field.

Next, we extended the functionality of Pt/LBMO to incorporate ferroelectric switching by a two-step process, as shown in Fig. 1c schematics with parameters described in *Methods*. The good sample quality can be inferred from the top-view optical micrograph (Fig. 1d), atomic structure from cross-sectional scanning transmission electron micrograph (STEM) (Fig. 3), and X-ray diffraction (XRD) (Supplementary Fig. 1b). In Fig. 1e, by connecting the bottom SRO and the top Hall-bar, good polarization-gate voltage ($P$-$V_{gate}$) loops can be obtained from the sandwiched PZT, showing remanence of ~75 μC/cm$^2$. The coercive voltage may increase slightly if the switching time is shortened.

Then, we measured the $\rho_{xy}$ from the top Hall-bars at various temperatures after applying $V_{gate}$ pulses of +/-3V of 1 millisecond width. Fig. 2a shows that obvious THE humps exist when the polarization is pointing down (***P*$_{DOWN}$**) away from LBMO but are suppressed at upward polarization (***P*$_{UP}$**), corresponding to Skyrmion creation and annihilation respectively. By fixing the temperature at 175 K magnetic field at 0.2 T where such change is the most obvious, $\rho_{xy}$ versus sweeping $V_{gate}$ unambiguously traces a hysteresis loop (Fig. 2b), proving that the Hall response mentioned is directly related to the PZT's ferroelectric polarization. From the absence of THE humps at $P_{UP}$, we may infer that the magnetic energy landscape is changed to become unfavourable for Skyrmion formation, i.e. the domain wall energy $\sigma_{DW} = 4\sqrt{A_{ex}K_{eff}} - \pi D_{int} > 0$, where $K_{eff} = K_U + \mu_o M_{sat}^2/2$ is the effective anisotropy with the dipolar interaction accounted, while $A_{ex}$, $K_U$ and $D_{int}$ and $M_{sat}$ are the exchange stiffness related to $T_C$, uniaxial anisotropy, interfacial DMI and saturation magnetization. Upon switching the ferroelectric polarization, modulation of all three parameters $A_{ex}$, $K_U$ and $D_{int}$ are possible, and we will attempt to pinpoint the more dominant tuning factor in the later part of this paper. Although the tuning efficacy reduces with increasing temperature, it is possible to retain a significant $\Delta\rho_{THE}$ up to room temperature by increasing the LBMO thickness slightly to 9uc, as shown in Fig. 2c.

In Figure 3, we show a detailed analysis on the cross-sectional atomic structure by STEM, where the sample was cut directly from the Hall bar's centre encompassing the Pt/LBMO/PZT/SRO//STO(001) by focused ion beam (FIB). The depth profiling of lattice parameters along all 3-axes shows strain relaxation mainly in the PZT layer. The c-axis lattice parameter values are also consistent to our 2theta-omega XRD scan around (002).

We now discuss the relevant physics upon switching the ferroelectric polarization of PZT. Firstly, from the design perspective, since no single-crystalline oxide can be grown on amorphous Pt, growing LBMO on PZT is necessary to form the interface with Pt. The resulted LBMO crystallinity is less than that of epitaxial LBMO//STO(001), as evident from a higher surface roughness of 0.3 nm extracted from topography scan and full-width-half-maximum (FWHM) at XRD ω-rocking curve of (002), suggesting a mild Stranski-Krastanov (SK) growth (supplementary Fig. S2). Hence, this factor would reduce the $M_{sat}$ of LBMO, if comparing the same thickness between that grown on PZT and STO(001). The valence electronic configuration in perovskite hole-doped manganites is well-known to be 3d $t_{2g}^3$ $e_g^{1-x}$ where x is the doping percentage, and all electrons are spin-up since the Hund's splitting is way higher than crystal-field splitting. Following the consensus of magnetoelectric coupling established at PZT/La$_{1-x}$Sr$_x$MnO$_3$(LSMO) interfaces[21], hole depletion is expected at polarization pointing towards LSMO, which reduces the Mn$^{(3+x)+}$ valence and promotes z-direction (interlayer) double-exchange (DE) via $3d_{3z^2-r^2}$ orbitals, thus increasing the $M_{sat}$. Vice versa, polarization pointing away from LSMO weakens the z-direction DE but promotes interlayer super-exchange (SE) instead, whereas the xy-direction DE remains strong, thus a lower $M_{sat}$ is resulted due to the presence of an A-type antiferromagnetic (A-AF) "dead-layer" near PZT/LSMO interface. Such phenomenon has been reproduced in several experiments with PZT-on-LSMO structures[43, 44] and Density Functional Theory (DFT) calculations[45, 46]. Exceptionally, reference [47] using LSMO/PZT/LSMO trilayer but without polarization switching obtained a comparison result opposite to the established consensus. Their analyses were likely plagued by the mentioned quality problem with LSMO-on-PZT creating an even thicker dead-layer in the top LSMO. However, in our case, the tuning of LBMO's magnetic properties is directly resulted from polarization switching, thus films' quality issue is excluded in comparing the effect of polarization direction, yet $P_{up}$ towards LBMO still results in weakened apparent Hall Effect, opposite to the consensus.

More crucially, our LBMO's c-axis lattice parameter of 4.04 Å is far larger than its bulk value 3.88 Å, as measured by XRD (supplementary Fig. 2a). This is opposite to the structures involved in the aforementioned consensus where the LSMO//STO(001) (below the PZT) is usually under in-plane tensile, c-axis compression by STO(001). Here, the crystal field in a c-axis elongated manganite unit cell mandates lower $3d_{3z^2-r^2}$ orbital energy compared to that of $3d_{x^2-y^2}$, hence $3d_{3z^2-r^2}$ and $3d_{x^2-y^2}$ are over- and under-populated by electrons, respectively. Then, polarization towards LBMO ($P_{UP}$) further redistributes or siphons electrons from $3d_{x^2-y^2}$ to the $3d_{3z^2-r^2}$ orbitals, promoting z-direction DE but xy-direction SE, creating a C-type antiferromagnetic (C-AF) dead layer. Vice versa, polarization away from LBMO ($P_{DOWN}$) would restore a more balanced DE among all directions, akin to cancelling the strain effect. This way, the $M_{sat}$ for $P_{DOWN}$ is stronger than that of $P_{UP}$. This concept is illustrated in Fig. 3a,b for the two polarization directions. The insight gained here is analogous to the A-AF found in LaMnO$_3$//SrTiO$_3$(001) but C-AF in LaMnO$_3$//NdGaO$_3$(110). In other words, the electrostatic doping and strain effects act in tandem in deciding the tuning polarity of interfaces between hole-doped manganite and PZT (or other $P_{[001]}$ ferroelectrics in general). Lastly, we provide micromagnetic simulation to illustrate the tuning effect in Fig. 3c,d. Since slightly weaker magnetic properties can be expected by growing LBMO on PZT, we arbitrarily reduced the magnitudes of $A_{ex}$, $M_{sat}$ and $K_U$ in the simulation into 2.5 pJ/m, 250 kA/m and 7 kJ/m$^3$ respectively as compared to the parameter set shown in supplementary Fig. S1, while ensuring the peak-TCD fields matching with the experimental THE result. Hence at the $P_{DOWN}$ ferromagnetic state with $A_{ex}$ = +2.5 pJ/m, a Skyrmion-lattice (SkL) can be obtained at $\mu_o H_z$ = +/-0.25 T (Fig. 3c). Then, as the sign of $A_{ex}$ is reversed (-2.5 pJ/m) at the $P_{UP}$ state to represent the C-AF phase, the resulted domain evolution becomes almost insensitive to the magnetic field, with small residual TCD values likely originating from the antiferromagnetic domain walls. Such concept would be unambiguous for the low temperature range (150 – 250 K), yet thermal fluctuation at higher temperatures and DMI weaken the C-AF phase rendering it indistinguishable from the ferromagnetic one and producing a noticeable THE as well.

In conclusions, the benefits gained from the device structure in this work are at least threefold. Firstly, the MPE of ultrathin Pt allows direct probing of magnetic properties via Hall Effect. Secondly, the magnetoelectric response of LBMO/PZT is huge upon ferroelectric polarization switching, resulting in creation/annihilation of magnetic Skyrmions and qualifying as a high-temperature artificial multiferroic Skyrmion-host. Thirdly, we completed the bigger picture about the competition between electrostatic doping and strain in tuning orbital occupancy and deciding the resulted magnetic phases, since our observation is novel yet complementary to the earlier understanding. The device structure also holds the potential for making Skyrmion field-effect transistors[48, 49], possibly integrating with the tunnelling non-collinear magnetoresistance (TNcMR)[50], to find application in building neuromorphic networks.

**Methods:**

All samples were grown in a pulsed laser deposition (PLD) system with in-situ transfer to a DC sputtering chamber. The bottom electrode (20 nm SRO) was first grown at 650°C 100 mTorr to achieve the typical step-flow quality and was subsequently defined *ex-situ* into long bars by photolithography and Argon ion milling (step 1). The subsequent LBMO/PZT layers were continued at 600°C to prevent the evaporation of SRO and promote better PZT's quality, followed by the deposition of 2nm amorphous Pt by DC sputtering at room temperature (step 2). The PZT/SRO interface remains clean and similar to in-situ interfaces due to the reheating, although vacuum has been broken at step 1. Finally, the top Hall-bars was defined by another photolithography and ion milling to the full depth of LBMO. The resulted leakage current between the top Hall-bars and the bottom electrodes is <1 nA.

Electrical measurements were performed in a Quantum Design Physical Measurement System (PPMS). A Stanford Research SR830 Lock-in Amplifier was used to measure the Hall Effect by sourcing an AC current of 15 μA to the sample connected in series to a standard 100 kohm resistor. A Keithley Pulse Sourcemeter Model 2430 was used to apply the DC $V_{gate}$ pulses during Hall measurements, while the $P$-$V_{gate}$ hysteresis loops were obtained from a Radiant Technologies Premier II Ferroelectric Tester.

X-ray diffraction was measured at a 4-circle high-resolution diffractometer located at the XDD beamline of Singapore Synchrotron Light Source (SSLS).

**Acknowledgments:**

This research is supported by the Agency for Science, Technology and Research (A*STAR) under its Advanced Manufacturing and Engineering (AME) Individual Research Grant (IRG) (A2083c0054) and by the National Research Foundation (NRF) of Singapore under its NRF-ISF joint program (Grant No. NRF2020-NRF-ISF004-3518).

**Declarations:**

The authors declare no competing interests.

**Main Text Figures:**

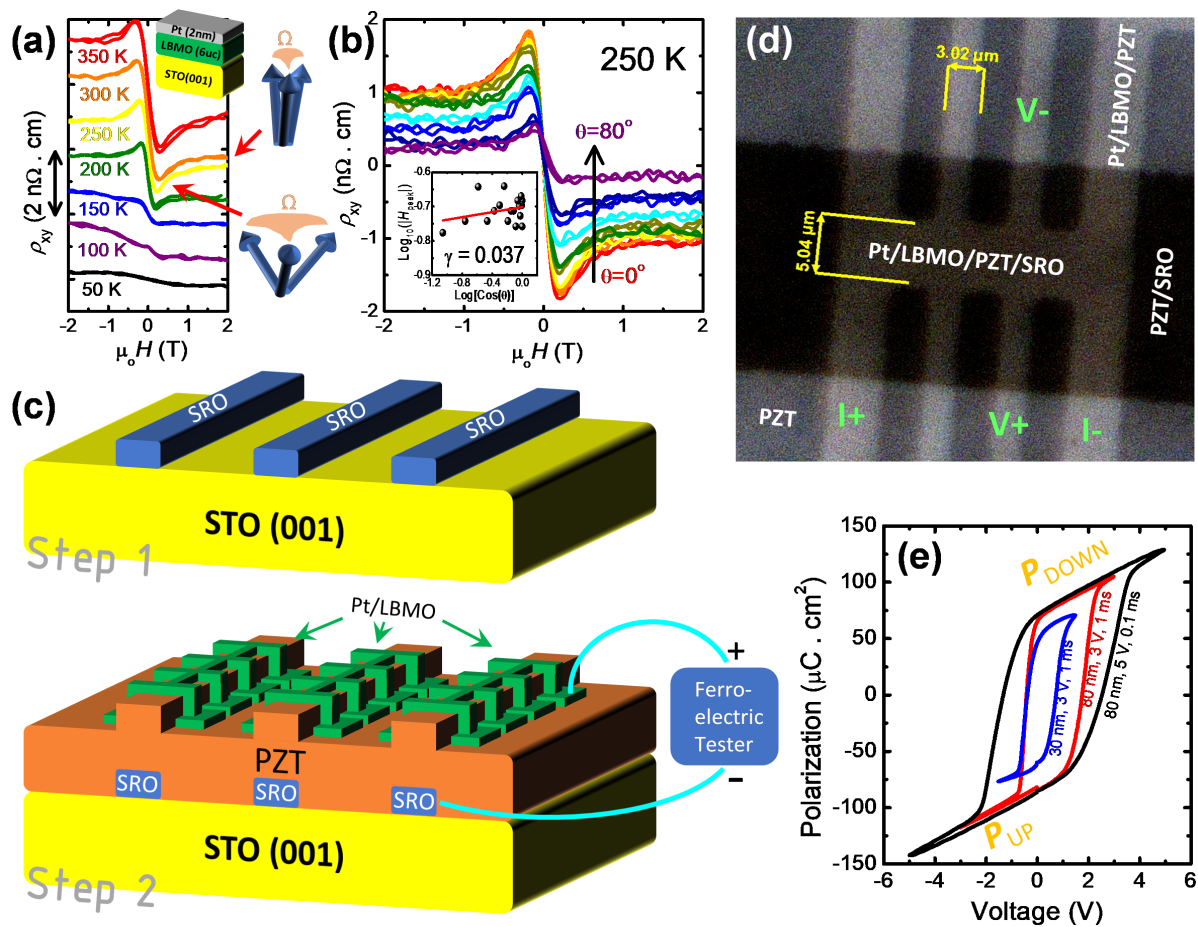

**Figure 1: Basic characterizations and device fabrications. (a)** Hall Effect of ultrathin Pt/LBMO bilayers, where the THE evolutions can be understood by the solid angles Ω subtended by magnetic moments as illustrated in the insets. **(b)** θ-dependent Hall Effect of Pt/LBMO for extraction of its (near-zero) γ-exponent, following the convention of $\mu_o H_{peak} \propto \cos^\gamma(\theta)$. **(c)** Schematics of Pt/LBMO/PZT/SRO device fabricated in a two-step process, and **(d)** its resulted top-view optical micrograph. **(e)** $P$-$V_{gate}$ loops measured across the PZT layer with varying PZT thickness and pulse widths.

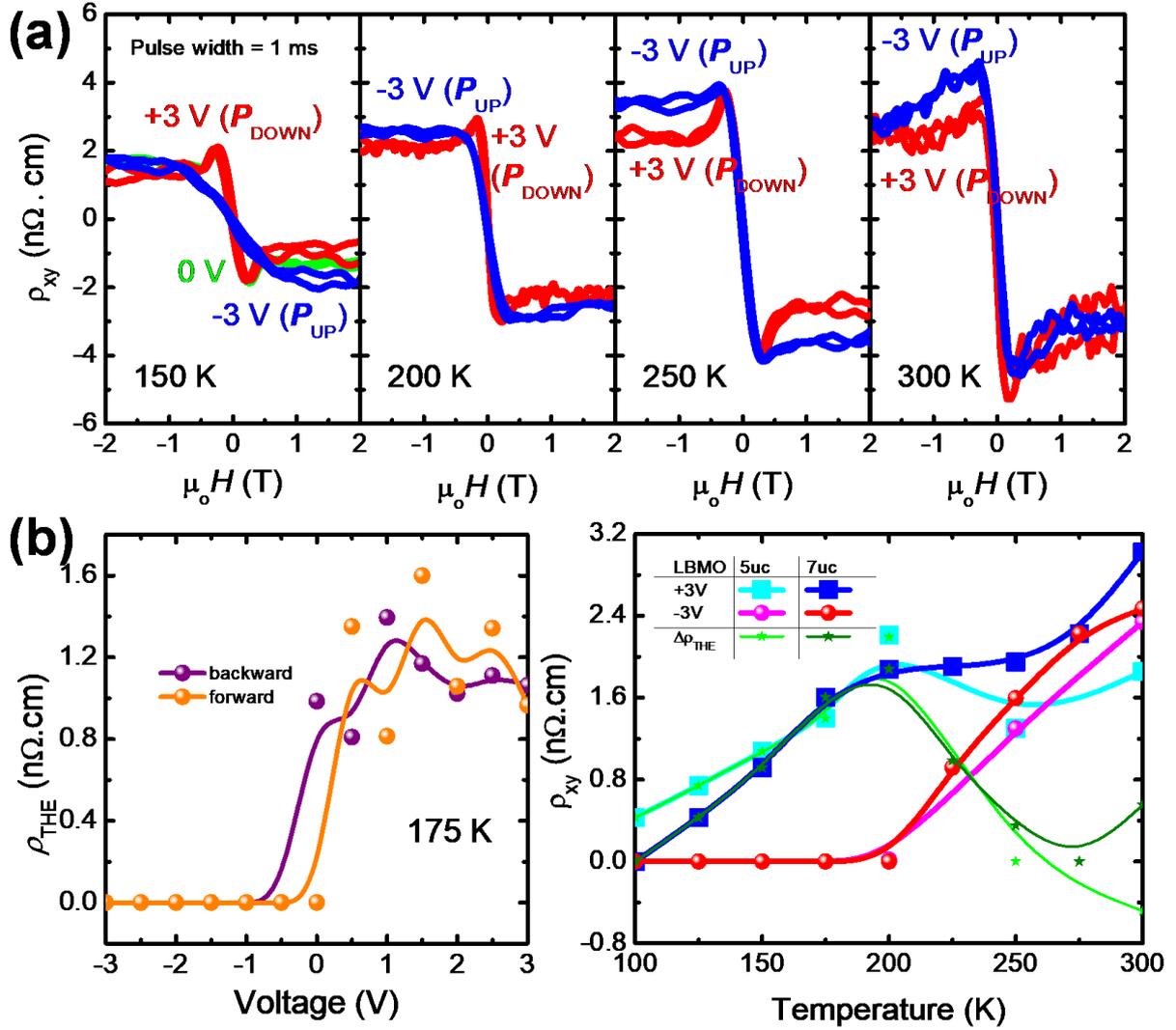

**Figure 2: Magnetoelectric effect of Pt/LBMO in response to switching PZT polarization.** (a) Hall Effect data measured at various temperatures after each $V_{gate}$ pulse of +/-3V of 1 ms width. (b) A particular temperature (175 K) was selected to perform the $V_{gate}$ sweep, where a hysteresis loop of THE resistivity versus $V_{gate}$ can be extracted. (c) Temperature-dependent THE resistivities measured after the application of the $V_{gate}$ = +/-3 V pulses, for varying LBMO thickness, showing the potential of extending the magnetoelectric tuning window to room temperature.

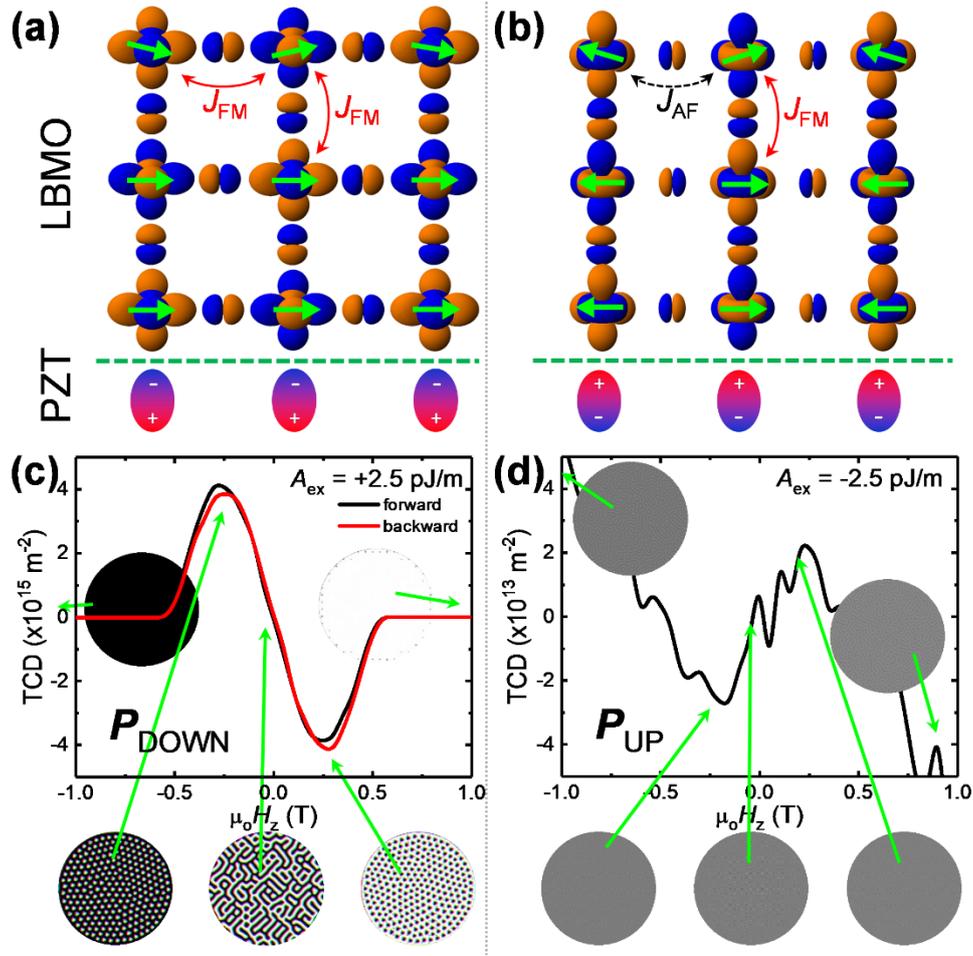

**Figure 3: Mechanism of the LBMO/PZT magnetoelectric response.** Illustrations showing the double-exchange ($J_{FM}$) and super-exchange ($J_{AF}$) pathways among $3d_{x^2-y^2}$ and $3d_{3z^2-r^2}$ orbitals (drawn overlapping each other) via the O2p orbitals, for the cases of polarization pointing **(a)** away from and **(b)** towards LBMO. The $t_{2g}$ orbitals containing $3\mu_B$ magnetization are not shown. Orbitals elongation or suppression are illustrated to show the electric field effect by the polarization. ferromagnetic (a) and C-type antiferromagnetic (b) moment arrangements are indicated by green arrows, while the progressive canting towards the top Pt/LBMO interface by the action of DMI is also illustrated. Plots of TCD($H_z$) obtained by MUMAX$^3$ simulations for the cases of polarization pointing **(c)** away from and **(d)** towards LBMO, where they differ by a positive (ferromagnetic) and negative (antiferromagnetic) signs in $A_{ex}$ respectively. The corresponding moment snapshots are also labelled at $\mu_o H_z = 0$ T, +/-0.2 T and +/-1.0 T, showing ferromagnetic textures evolution from stripes → SkL → saturation with increasing magnetic field in (c), but antiferromagnetic domain wall textures in (d).

**Supplementary Information:**

**A Robust High-temperature Multiferroic Device with Tunable Topological Hall Effect**

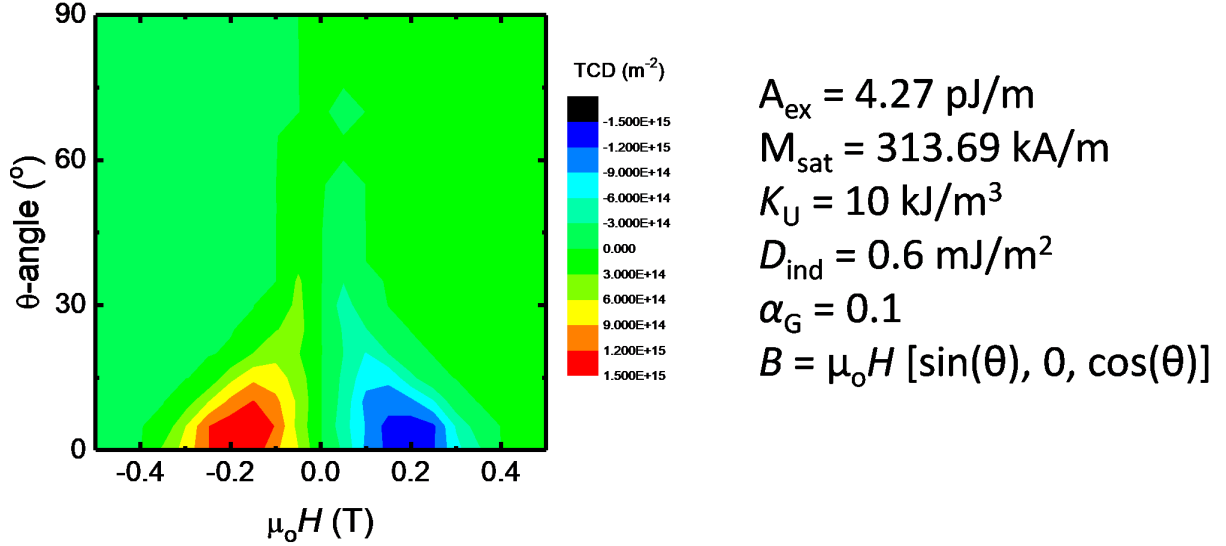

**Supplementary Figure 1:** Mapping of TCD versus total magnetic field and θ-rotation, showing the non-diverging behaviour of peak-TCD fields and Skyrmions' phase space, and supportive of main text Fig. 1b. The simulation parameters are shown in the right panel. Here, estimations follow: $A_{ex} = \frac{3k_B T_C}{2j(j+1)a}$ where $T_C$ is assumed to be ~300 K, j=3/2 due to the at least 3 $\mu_B$/Mn, and lattice parameter $a$ = 3.88 Å. $M_{sat}$ at zero Kelvin = $3\mu_B/a^3$ = 476.3 kA/m. With the mean field theory extension to the Langevin's paramagnetism, the temperature dependence of magnetization typically follows $M_{sat}(T)/M_{sat}(0\ K) = \tanh\left(\frac{M_{sat}(T)/M_{sat}(0\ K)}{T/T_C}\right)$, hence yielding $M_{sat}$=313.69 kA/m at 250 K. $K_U$ and intercial DMI ($D_{ind}$) are arbitrarily adjusted to match the peak TCD field with the experimentally measured peak THE fields at rotation angle θ=0. Swapping the x- and y-axes in the magnetic field vector into $\boldsymbol{B} = \mu_o H [0, \sin(\theta), \cos(\theta)]$ will produce the same result.

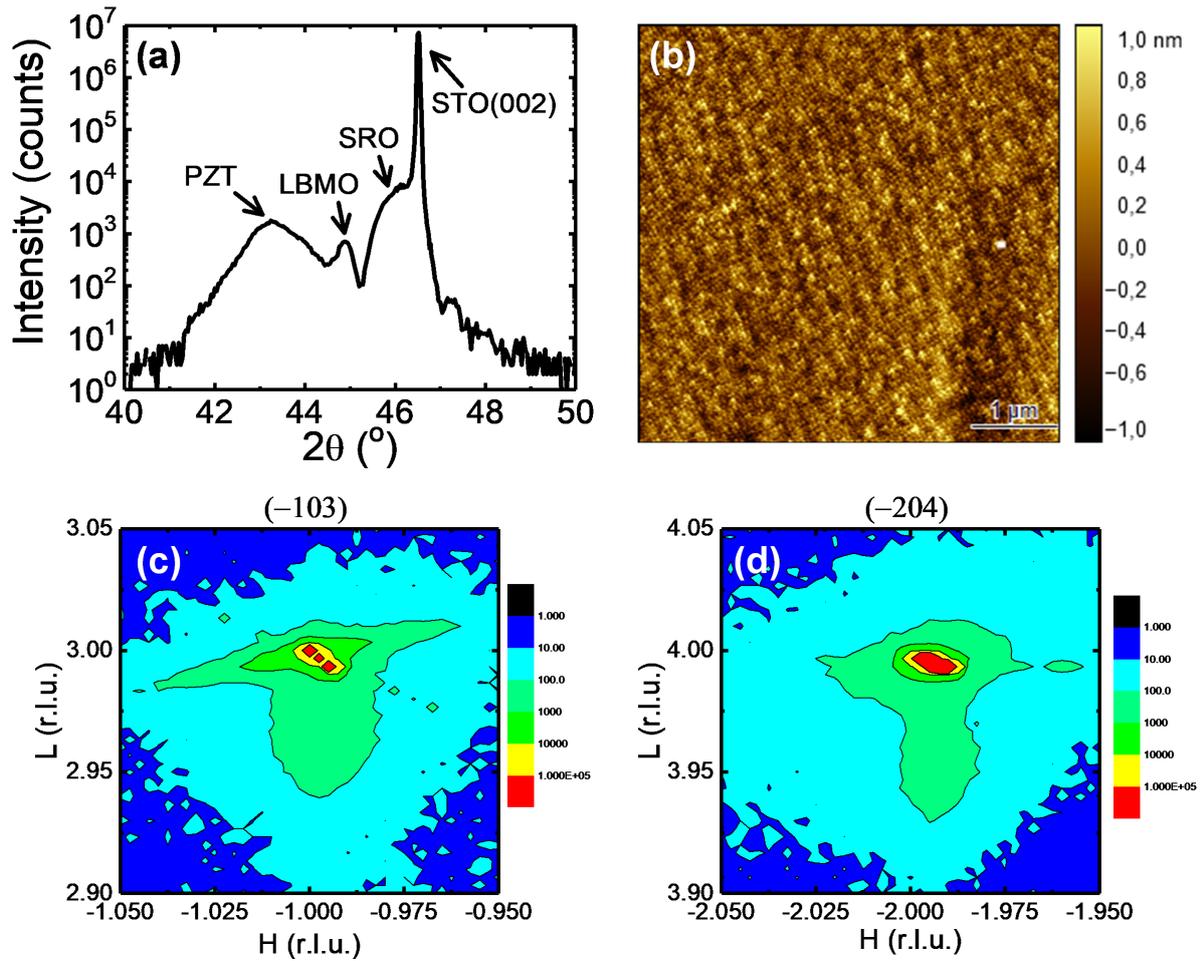

**Supplementary Figure 2: (a)** 2theta-omega XRD scan around (002) of the Pt/LBMO/PZT/SRO//STO(001) sample without ion milling. The peak positions are at 43.24°, 44.88°, 46.12° and 46.51° measured at a wavelength of 1.5419 nm at the SSLS XDD beamline, yielding c-axis lattice parameters of 4.1841 Å, 4.0394 Å, 3.9364 Å, and 3.9053 Å (standard) for PZT, LBMO, SRO, STO respectively. **(b)** Topography scan of the sample showing surface roughness of 0.31 nm, performed by an atomic force microscope using cantilever with force constant of ~15 N/m and resonant frequency of ~280 kHz. **(c)** and **(d)** XRD reciprocal space mapping of the sample for the (-103) and (-204) crystallographic planes, respectively.